\documentclass[preprint,3p,times]{elsarticle}

\usepackage{amssymb}

\usepackage{subfigure}
\usepackage{url}
\usepackage{float}

\usepackage{amsmath}
\usepackage{booktabs}
\journal{Elsevier}

\begin{document}

\begin{frontmatter}



\title{Can We ``Sense'' the Call of The Ocean?\\ Current Advances in Remote Sensing Computational Imaging for\\ Marine Debris Monitoring}


\author[1]{Oktay Karaku\c{s}}

\address[1]{Cardiff University, School of Computer Science and Informatics, Abacws, Senghennydd Road, Cardiff, CF24 4AG, UK}

\begin{abstract}
Especially due to the unconscious use of petroleum products, the ocean faces a potential danger: \textit{plastic pollution}. Plastic pollutes not only the ocean but also directly the air and foods whilst endangering the ocean wild-life due to the ingestion and entanglements. Especially, during the last decade, public initiatives and academic institutions have spent an enormous time on finding possible solutions to marine plastic pollution. Remote sensing imagery sits in a crucial place for these efforts since it provides highly informative earth observation products. Despite this, detection, and monitoring of the marine environment in the context of plastic pollution is still in its early stages and the current technology offers possible important development for the computational efforts. This paper contributes to the literature with a thorough and rich review and aims to highlight notable literature milestones in marine debris monitoring applications by promoting the computational imaging methodology behind these approaches. 

\end{abstract}




\end{frontmatter}


\section{Introduction}
\label{sec:intro}
The ocean provides a livelihood for more than 3 billion people whilst being the habitat for billions of species and generates more than \$3 trillion each year to the global economy \cite{worldEF}. Despite this importance, especially in the last decade, plastic pollution and marine debris have become one of the most important reasons for endangering this valuable and consumable resource due to the large needs of the human population and the unconscious use of plastic. 

\begin{figure}[t]
    \centering
    \includegraphics[width=\linewidth]{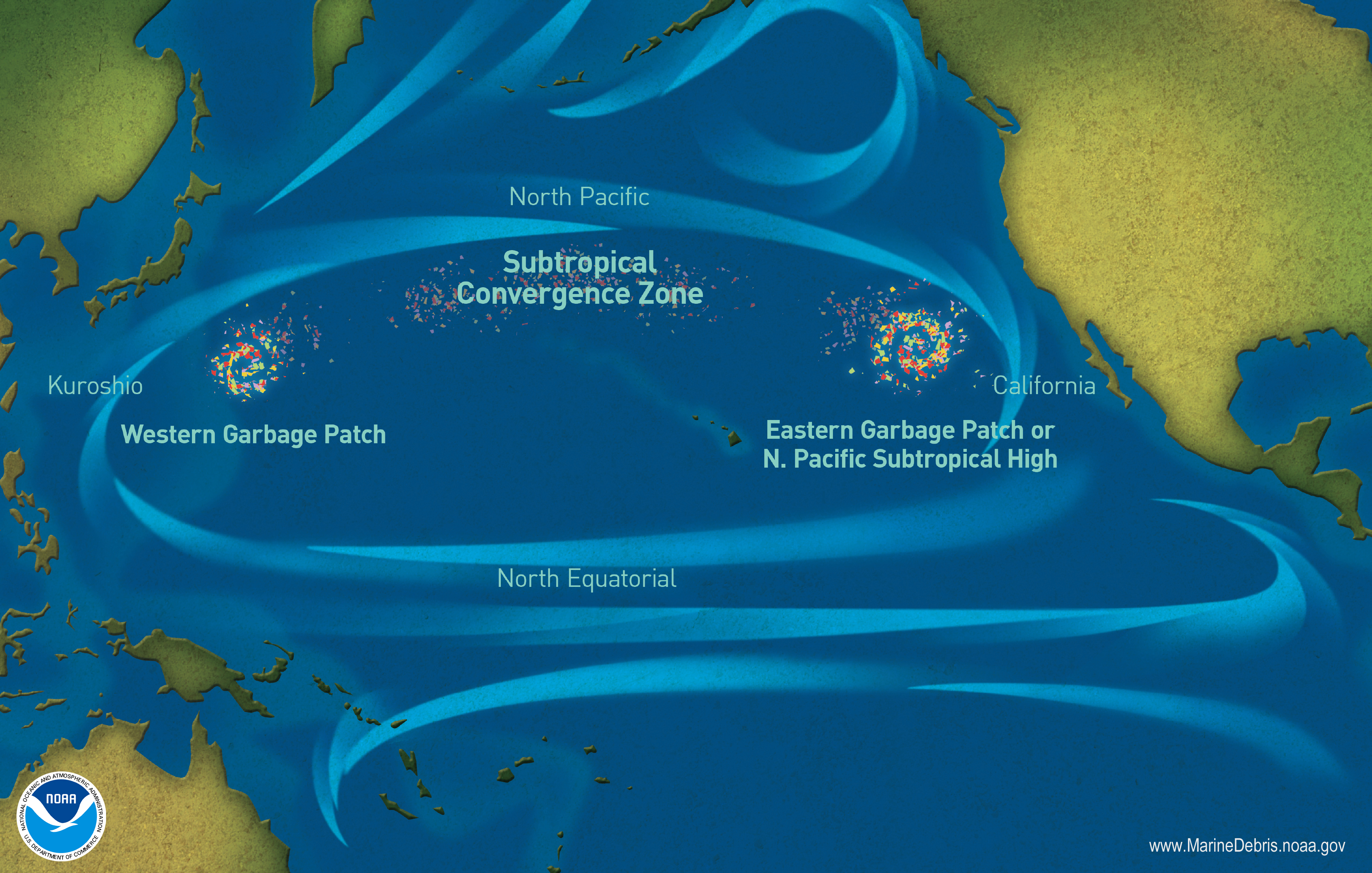}
    \caption[Great Pacific Garbage Patches]{Great Pacific Garbage Patches\footnotemark[1].}
    \label{fig:debris}
\end{figure}

Particularly, large plastics entering the ocean either sinks or floats since they are low-density materials \cite{farre2020remote}. Floating plastics on the macro-scale (say macro-plastics, $>$ 5mm) is the most crucial sub-group of floating plastics that cause entangling and ingestion for marine wildlife. Thus, detecting and tracking macro-scale floating plastics before their fragmentation into micro and nano scales, and/or threatening the wild-life through entanglement and ingestion becomes the primary action to protect the marine environment. One example area of focus is the Great Pacific Garbage Patch (GPGP), which can be described as a gyre within a gyre. GPGP covers an area of the ocean of approximately 1.6 million square km (roughly 8, 3, and 2 times greater than the UK, France, and Turkey, respectively) in size. However, the density of ocean plastic can vary greatly and change quickly within any location \cite{lebreton2018evidence}. Figure \ref{fig:debris} shows garbage patches and ocean currents in the Pacific Ocean. 

\begin{figure}[t]
    \centering
    \includegraphics[width=\linewidth]{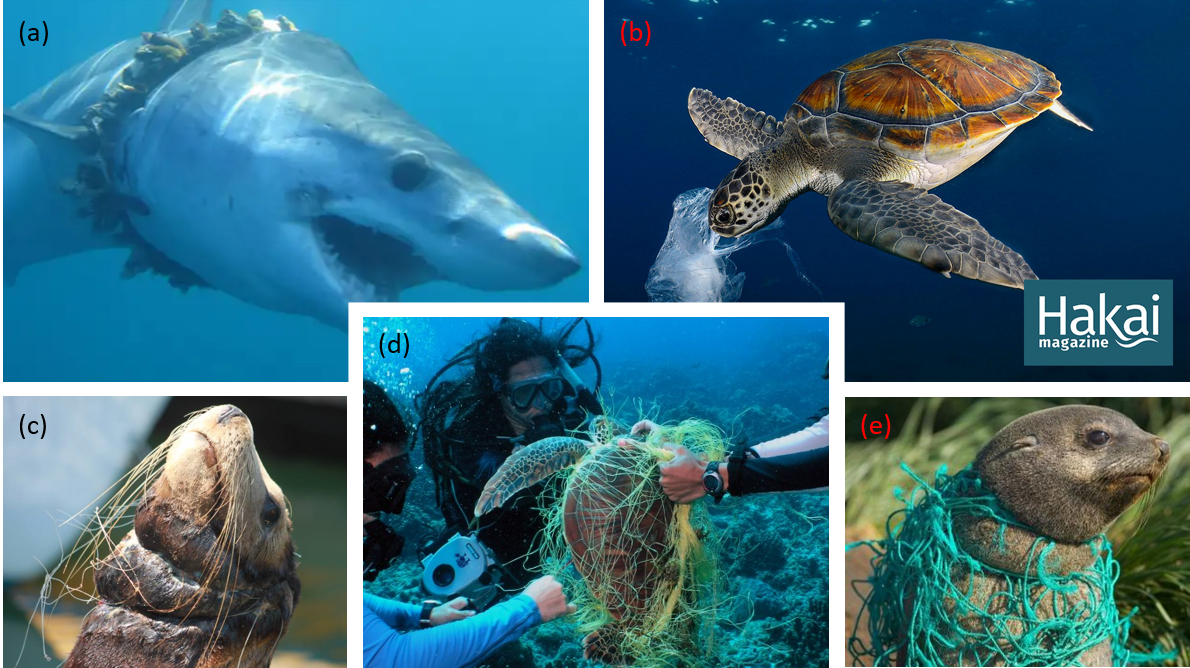}
    \caption[Plastic pollution threat on sea animals. (a) Firepaw, Inc., (b) Hakai Magazine, (c) Lauren Packard, (d) World Animal Protection, Marine Animals of Maine, and David Burdick-Marine Photobank, and (e) Planet Love Life.]{Plastic pollution threat on sea animals. (a) Firepaw, Inc.\footnotemark[2], (b) Hakai Magazine\footnotemark[3], (c) Lauren Packard\footnotemark[4], (d) World Animal Protection, Marine Animals of Maine, and David Burdick-Marine Photobank\footnotemark[5], and (e) Planet Love Life\footnotemark[6].}
    \label{fig:animals}
\end{figure}

Since the means of plastic usage, specifically in the last decade, has shifted from a positive tool to a potential danger to the environment, various efforts have been performed to help authorities and the academic community in reducing and preventing marine pollution. Marine debris and floating plastics can cause various problems, but mostly affects human health and marine wildlife. Particularly, plastic is polluting not only the water, but also the air we breathe, and the food we eat. According to the WWF's report in 2019, an average person could be ingesting approximately 5 grams of plastic every week, which is the equivalent of a credit card’s worth of microplastics. Moreover, an average person potentially consumes as much as 1769 particles of plastic every week just from water \cite{twoOceansAqu2020}. Apart from the risks it causes to human health, marine litter and debris also endanger wildlife crucially, from tiny coral reefs to huge whales. To date, plastic ingestion and/or entanglement has caused the killing of more than 700 marine species \cite{wwf2019plastic}. According to the research, more than 50\% of sea turtles eat plastic whilst plastic materials have been found inside more than half of the dead sea turtles. Additionally, more than 50\% of the sea mammals such as dolphins, whales and seals are impacted by plastic pollution whilst 100 thousand of them are killed by plastic each year (National Oceanic and Atmospheric Administration (NOAA) via \cite{wwf2019plastic2}). Figure \ref{fig:animals} depicts several examples of plastic pollution threat on sea animals.
\footnotetext[1]{\url{https://marinedebris.noaa.gov/info/patch.html}}

In addition to the problems caused for human health and marine wildlife, the marine debris problem causes some non-native organisms to travel to areas where they would not be found in normal circumstances. These organisms attach themselves to marine debris and are transported all around the world which then gain the capability to establish new neopelagic ecosystems \cite{haram2021emergence}. Last but unfortunately not least, the ocean pollution problem can heavily affect economies, especially for countries whose economy relies on summer tourism. Considering the summer tourism sector heavily relies on healthy coastal and ocean resources along with environmental and aesthetic quality, beaches where floating plastics and garbage patches are drifted, will not provide healthy conditions and this sector will economically be affected by this pollution problem \cite{noaadebris}.



For environmental applications, exploiting earth observation systems and their products such as active and passive imagery is a long-established approach, whilst for marine debris and floating plastic monitoring the efforts are still in the early stages. Airborne and space-borne remote sensing modalities such as the optical, hyperspectral, and synthetic aperture radar (SAR) have shown an important potential up to date in detecting, tracking, and mapping marine pollutants, such as oil spills. Moreover, thanks to their (optical, hyper-spectral) rich visible band and spectral choices, remote sensing imagery has had applications in assessing various metrics based on algal blooms whilst SAR has shown to be the gold standard in vessel-tracking and recognition applications. For all the marine applications above, the main imaging modality limitation was their spectral and temporal resolutions. Thanks to recent technological developments, new generations of satellites have been launched and spatial resolutions which were previously unavailable are now provided from space-borne remote sensing. In the last decade, initiatives such as NASA, UK Space Agency (UKSA), the European Space Agency (ESA), the International Ocean-Colour Coordinating Group (IOCCG), the Scientific Committee on Oceanic Research (SCOR), The Joint Group of Experts on the Scientific Aspects of Marine Environmental Protection (GESAMP) pay attention to saving the ocean by using recent technological advances. For this purpose, some effort has been made to track/detect marine pollution and illegal fishing. Satellite remote sensing imagery (optical, SAR, hyper-spectral, etc.) has been the leading technique for marine monitoring since its capability to capture wide-angle area imagery under challenging weather conditions and for various sea states, etc.
\footnotetext[2]{\url{https://firepaw.org/2019/07/05/sharks-other-marine-life-strangled-by-plastic-study/}}
\footnotetext[3]{\url{https://www.hakaimagazine.com/news/plastic-pollutions-rapidly-mounting-toll/}}
\footnotetext[4]{\url{https://www.flickr.com/photos/110485367@N08/11180067935}}
\footnotetext[5]{\url{https://www.worldanimalprotection.org/our-work/animals-wild/sea-change/our-work/rescue}}
\footnotetext[6]{\url{https://www.planetlovelife.com/pages/wildlife-entanglement}}
\footnotetext[7]{\url{https://www.plastic-i.com/}}
\footnotetext[8]{\url{https://theoceancleanup.com/}}
\footnotetext[9]{\url{https://theoceancleanup.com/plastic-tracker/}}

Apart from the governmental initiatives mentioned above, various efforts have also been specifically spent on marine debris and floating plastics monitoring in both academia and industrial companies. Currently, creators of Plastic-i\footnotemark[7] have developed a powerful tool (and also the world's first) for mapping ocean plastics via satellites, and won the first-ever Hack The Planet Competition 2021. Moreover. A non-profit organisation, the Ocean Cleanup\footnotemark[8], is developing and scaling technologies to rid the oceans of plastic. Their software - The Plastic Tracker\footnotemark[9] - is an educative tool to predict where the floating plastic patches are heading to. Debris Tracker\footnotemark[10] is designed to make community members contribute data on plastic pollution. Passionate people from all around the world record data on inland and marine debris by using the Debris Tracking App. Ocean Scan\footnotemark[11] is a global database that aims to promote important international collaborations for marine debris cleaning efforts. They offer access to various in-situ observations and matching remote sensing images of marine litter and debris. Messages in Bottles (MiBs)\footnotemark[12] provides a non-real-time interactive tool to visualise information on coastal plastic waste using Sentinel-2 optical imagery (inspired by \cite{biermann2020finding}) in order to connect research and public interests for the purposes of the marine pollution efforts. Project Kaisei\footnotemark[13] is another clean-up initiative of Ocean Voyages Institute\footnotemark[13], a non-profit organization founded in 1979 the main focus of which is on major ocean clean-up and raising awareness regarding the global problem of marine debris/ocean trash. The International Coastal Cleanup (ICC)\footnotemark[14], with the help of volunteers from all around the world, dedicates itself to removing trash from the world’s beaches and waterways.
\footnotetext[10]{\url{https://debristracker.org/}}
\footnotetext[11]{\url{https://www.oceanscan.org/}}
\footnotetext[12]{\url{https://messages-in-bottles.netlify.app/}}
\footnotetext[13]{\url{https://www.oceanvoyagesinstitute.org/project-kaisei/}}
\footnotetext[14]{\url{https://oceanconservancy.org/trash-free-seas/international-coastal-cleanup//}}

Within this developing application area, the machine/deep learning-based remote sensing computational imaging approaches are still in the early stages one of the most important reasons for which is the lack of approved marine debris data sets. Especially in the last couple of years, there has been a good amount of academic outcomes published promoting advanced computational imaging approaches and publishing data sets thanks to the efforts from academic institutions for marine debris and floating plastics detection/tracking purposes. 

This paper, via contributing to the literature with a thorough review, highlights current advances based on remote sensing computational imaging for marine debris and floating plastics detection and tracking applications. The rest of the paper is organised as follows: Academic outcomes serving the efforts of marine debris problem are discussed in Section \ref{sec:pollution}. The literature statistics describing the number of publications, etc. is presented in Section \ref{sec:stats} whilst research challenges and future research directions are discussed in Sections \ref{sec:chal}, and \ref{sec:future}, respectively. Section \ref{sec:conc} concludes the paper with a general discussion.

\section{Marine Debris Monitoring Applications}\label{sec:pollution}
Initial efforts for the purposes of detecting, analysing the marine debris problem mostly rely on in-situ approaches via (1) visual surveys \cite{thiel2011spatio,lavers2017exceptional}, and/or (2) airborne data owing to their potential to monitor by the help of high-resolution and real-time information \cite{garaba2018airborne, moy2018mapping, themistocleous2020investigating}. In a recent study \cite{de2021quantifying}, de Vries et. al. present the utilisation of footage of floating plastic debris recorded offshore with GPS-enabled action cameras aboard vessels. Their work provides consistent results of predicted macroplastic compared to the global plastic dispersal models.

Due to its capability in covering large and inaccessible areas, satellite-based remote sensing imagery is considered a useful tool for monitoring the marine environment and detecting/tracking macro-scale marine pollution. An initial study promoting utilisation of satellite imagery in order to detect floating plastics/debris is by Aoyama \cite{aoyama2014monitoring,aoyama2016extraction}. This study firstly concludes that the small size of most marine debris means that it cannot be confirmed directly, even when using high spatial resolution satellite imagery. In order to solve these problems, the author proposed using several candidate debris pixels and making a spectral analysis on these whether there is a discrimination between debris and surrounding ocean pixels. Their approach works effectively in cases where the size or area of marine debris is large enough to affect a change in the captured spectra. Multi-spectral satellite imagery has also been used to validate various important post-events after the devastating tsunami triggered by the Tohoku-Oki earthquake of 11 March 2011. In \cite{matthews2017dynamics}, it has been shown that high spatial resolution satellite tracking reveals faster-than-expected floating debris motions.

A study by Goddijn-Murphy et al. \cite{goddijn2018concept} has shown that plastic litter and seawater develop a reflectance model coming from their spectral signatures and optical geometry. The work in \cite{goddijn2018concept} considers only one type of macro plastic and proposes that the fraction of a plastic surface can be estimated from the surface reflectance, since specifically knowing the clear water reflectance. It has also been reported in \cite{goddijn2018concept} that there are some limitations to this approach such as shading of surface plastic, and the key approach is to select a spectral wavelength in which water leaving light is minor and plastic reflectance is high (approx. 750 $\eta$m). Even though it leads future studies on plastic detection applications, the work in \cite{goddijn2018concept} hence constitutes limitations in natural environments since plastic debris appears in a magnitude of shapes, colours, and sizes. Nevertheless, determining the appropriate reflectance for the plastic makes it detectable in remote sensing imagery. 

\begin{figure}[ht]
    \centering
    \subfigure[]{\includegraphics[width=0.54\linewidth,keepaspectratio=true]{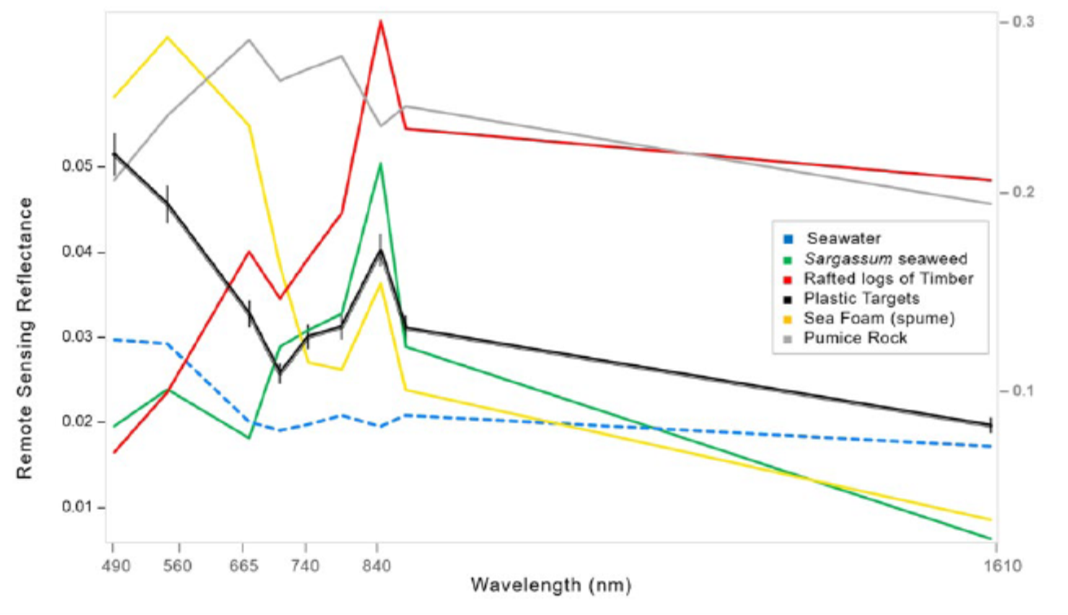}}
    \subfigure[]{\includegraphics[width=0.45\linewidth,keepaspectratio=true]{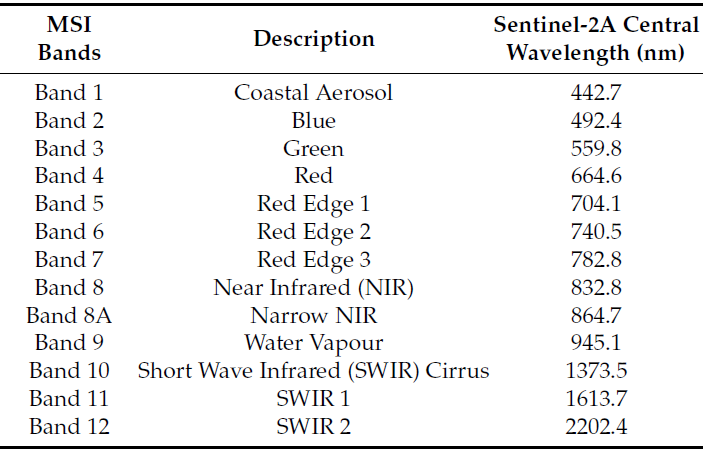}}
    \subfigure[]{\includegraphics[width=0.7\linewidth,keepaspectratio=true]{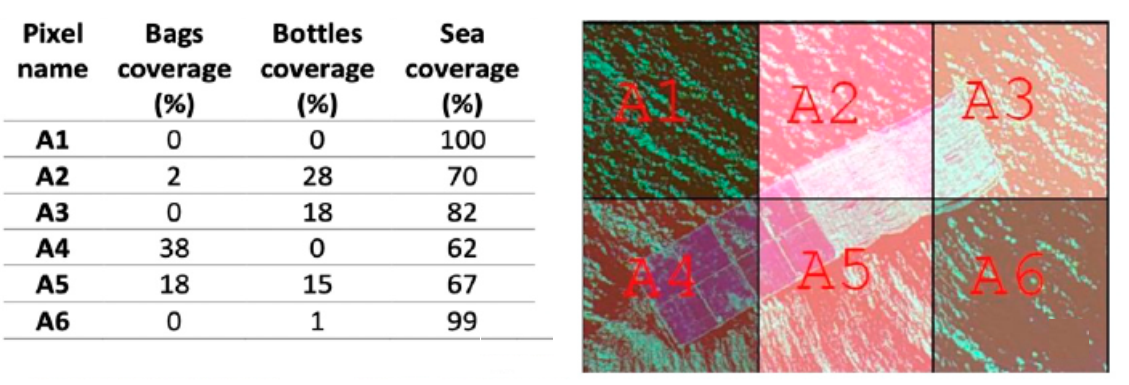}}
    \caption{(a) Spectral signatures of different materials \cite{biermann2020finding}. (b) Sentinel-2 band characteristics. (c) Example plastic coverage in pixels \cite{topouzelis2020remote}}
    \label{fig:sen2_1}

\end{figure}

Reflectance analysis in \cite{goddijn2018concept} motivated \cite{topouzelis2019detection,topouzelis2020remote} for a further detailed analysis via an approach consisting of both satellite and unmanned aerial imagery for the detection of plastic patches. Topouzelis et. al. \cite{topouzelis2019detection} have created a measurement setup at Tsamakia beach, Greece, which includes a set of three artificial floating plastic targets (PET bottles, plastic bags and fishing ghost nets). It has been shown that plastic litter such as bottles, bags, and fishing nets, reflect light in the near-infrared (NIR) band where clear water absorbs the light. Moreover, it has been reported in \cite{topouzelis2019detection} that the reflectance intensity is also directly related to the amount of plastic in a single pixel of the utilised image (10m x 10m for Sentinel-2). Thus, if water composes more than 50-70\% for a given pixel, the reflection from the plastic pollutant is relatively problematic in the NIR band. Consequently, remote sensing data is required to have pixels that are filled with more than 30-50\% plastics (an area of at least 30-50 m2). This case can be achieved for the regions of such as coastal areas, where an eddy or a similar reason might combine plastics into larger patches.  However, especially for the ocean, individual plastic patches will potentially be less than the detectable limit \cite{biermann2020finding}. This can be seen as the main limiting factor of the plastic detection research via the optical remote sensing modalities since the Sentinel-2 imagery has a spatial resolution of 10x10 meters. It leads to the detectable floating plastic area to be at least 50m$^2$ since plastic pollutants become spectrally visible if the water composes less than 50\% of the pixel area. In \cite{topouzelis2019detection}, a Sentinel-1 SAR imagery has also been analysed for the same measurement setup, however, it has not been concluded notable results based on the detection of plastics via SAR imagery. The potential main reason behind this is the spatial resolution of 10m and the relatively less penetrable frequency band of the Sentinel-1 satellites (C band compared to L or S bands).

Another work \cite{martinez2019measuring} has deeply discussed the requirements of a specifically-designed remote sensing monitoring system of plastic pollution. It has been reported in \cite{martinez2019measuring} that an ideal remote sensing system would compromise both passive (optical) and active (SAR) satellite modalities, as well as support from UAV based systems. Passive sensors that offer short-wave infrared (SWIR) band measurements are noted as the high potential modality, whilst SAR also specified as a potential for future developments, especially thanks to new SAR platforms (such as the ICEYE) capability to offer a higher spatial resolution, and lower revisiting times.

\begin{figure}[htbp]
    \centering
    \includegraphics[width=0.75\linewidth,keepaspectratio=true]{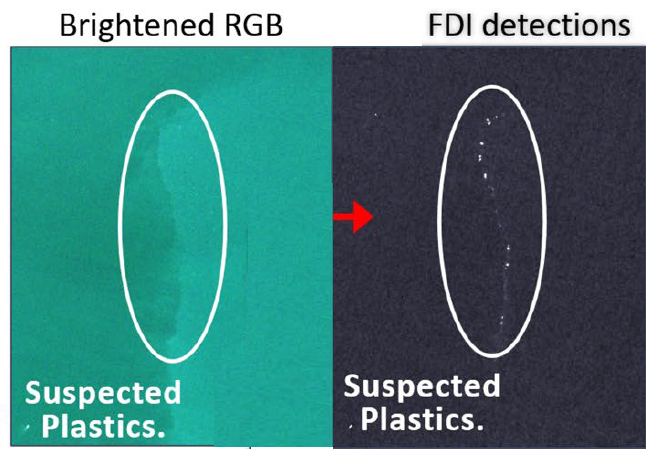}
    \caption{Representation of FDI index in optical imagery \cite{biermann2020finding}}
    \label{fig:fdi}
\end{figure}

The very first work on detecting and classifying plastic patches using \emph{solely} optical satellite data has been studied by Biermann et. al in \cite{biermann2020finding}. The work in \cite{biermann2020finding} has two joint objectives: (1)  demonstrating the capability of Sentinel-2 data on detecting floating macroplastics, and (2) classifying macroplastics and other natural materials which are most likely to be aggregated within mixed patches of floating debris. The authors proposed a novel parameter, which is the floating debris index (FDI), in order to analyse sub-pixel interactions of macroplastics with the sea surface to increase the chance of detection of patches floating on the ocean surface. The FDI index is inspired by a previously proposed index of floating algae index (FAI). By replacing the red band measurements with the red edge band (approx. at 740 nm), the authors leverage the difference between the NIR band and its baseline reflectance. The FDI can be represented as
\begin{align}
    FDI &= R_{NIR} - R'_{NIR}\\
    R'_{NIR} &= R_{RE2} + (R_{SWIR1} - R_{RE2})\times \dfrac{\lambda_{NIR} - \lambda_{RED}}{\lambda_{SWIR1} - \lambda_{RED}}\times 10
\end{align}
where $R_{NIR}, R_{RE2}$, and $R_{SWIR1}$ refers to NIR, red edge 2 and SWIR1 bands, respectively whilst similarly $\lambda$ values are wavelength values (in $\eta$m). Sentinel-2 band characteristics are provided in Figure \ref{fig:sen2_1}-(b). The FDI dramatically highlights the plastic and has been found useful in the identification of floating plastics in waterbodies using spectral remote sensing imagery. An example for the visibility of the FDI is depicted in Figure \ref{fig:fdi}.

It has also been shown in \cite{biermann2020finding} that using FDI in conjunction with the Normalised Vegetation Difference Index (NDVI) \begin{align}
    NDVI = \dfrac{R_{NIR} - R_{RED}}{R_{NIR} + R_{RED}}
\end{align}
makes it possible to detect differences between plastics, vegetation, driftwood, and seafoam.

Themistocleous et al. \cite{themistocleous2020investigating} set up a pilot study to investigate whether plastic targets on the sea surface can be detected from Sentinel-2 only data. Their setup consists of a target of plastic water bottles with size of 3m$\times$10m in the sea near the Old Port in Limassol, Cyprus. They gathered UAV multi-spectral images during the same time as the Sentinel-2 satellite passes. They have used various important spectral indices such as NDVI, but not using FDI of \cite{biermann2020finding} since the work has been published during the same time interval in 2020. Instead, they proposed a new index called Plastic Index (PI)
\begin{align}
    PI = \dfrac{R_{NIR}}{R_{NIR} + R_{RED}}.
\end{align}
Their spectral analysis on the visibility and detectability of the floating plastics in Sentinel-2 imagery showed that the newly developed PI has been able to identify floating plastics and was selected as the most effective index.

\begin{figure}[ht]
    \centering
    \includegraphics[width=\linewidth]{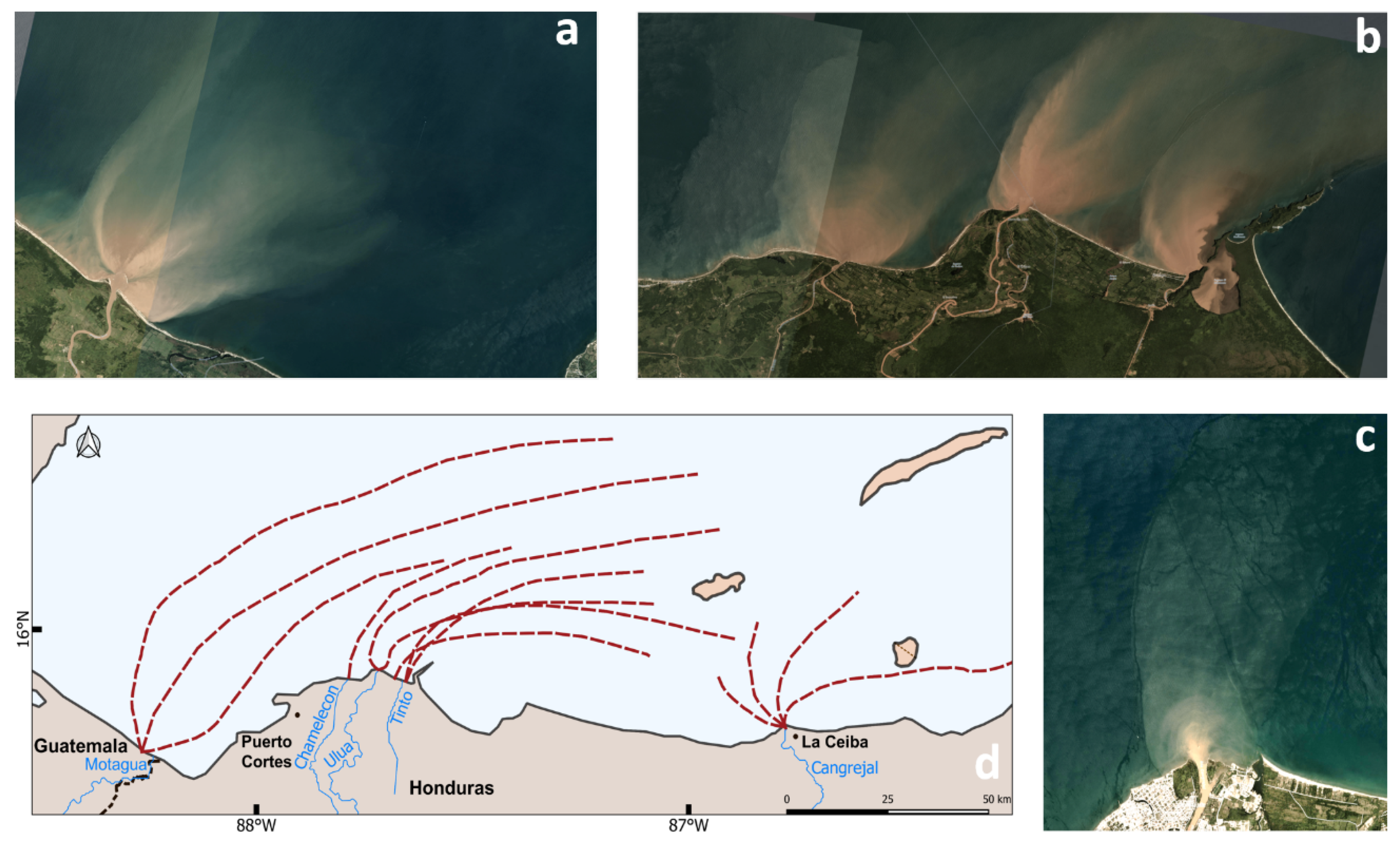}
    \caption{An example figure from \cite{kikaki2020remotely} shows the source of plastic pollution in Bay Islands in the Caribbean Sea.}
    \label{fig:kikaki}

\end{figure}

Kikaki et. al. \cite{kikaki2020remotely} has conducted a research over a specific target location - Bay Islands in the Caribbean Sea where remarkable amounts of plastic debris have been reported. Satellite imagery for the time interval of 2014-2019 has been investigated and in-situ data collected with vessel and diving expeditions has been verified via exploiting the spectral characteristics of the floating plastics. This work, even though it is not proposing a technical method to detect and track marine debris, utilises classical spectral analysis approaches, but this time to analyse the source of pollution. According to their findings, the main source of pollution in the target area has been found to be the river discharges from the basins of Honduras and Guatemala. It has also been reported that dynamic sea currents affects the pollution and plastic patches can travel more than 200km. Figure \ref{fig:kikaki} shows an example figure highlighting the source of plastic pollution in Bay Islands in the Caribbean Sea.

Up to this point, all the works were based on Sentinel-2 imagery for the purpose of promoting satellite images in floating plastic detection. Park et. al. \cite{park2021detecting}, instead of Sentinel-2 multi-spectral imagery, utilises very high geo-spatial resolution 8-waveband WorldView-3 imagery in order to observe floating plastic litter in the Great Pacific Garbage Patch (GPGP). They apply various spectral analysis approaches and investigate anomalies to infer presence of
suspected plastic litter.

Salgado-Hernanz et. al., in their bibliographic review \cite{salgado2021assessment}, reveal one more time that the optical sensors seem to be the most optimum to target marine litter. Interestingly, in their work, they also promote SAR utilisation and specifically mention that SAR sensors may detect sea slicks (areas potentially containing high concentrations of microplastics). It is also important to note that depending on time scales and the background sea slicks may not contain marine debris. Furthermore, Hu \cite{hu2021remote} also suggests that it is impossible to detect floating material by all the optical spectra and analysis should be performed over the difference spectra to minimize the impact of variable subpixel
coverage. It is certain that some spectral bands are more important than others for the remote detection of marine debris. Ciappa \cite{ciappa2022marine} proposes marine litter detection work for the target location of the North Adriatic during the summer of 2020. This work similar to the ones mentioned above uses spectra arithmetic for spectral anomalies of the Red Edge bands, assuming changes of the red edge in pixels where marine litter was mixed with vegetal materials. Table \ref{tab:indices} highlights the most common spectral indices utilised by the works discussed above.

Knaeps et. al.\cite{knaeps2021hyperspectral} proposed literature with a data set of 47 hyperspectral-reflectance measurements of plastic litter samples in dry and wet conditions from the Port of Antwerp. Their results specifically highlighted water absorption and suspended sediments which could allow future research to appropriately select wavelengths. Garaba et. al.\cite{garaba2021concentration} proposed an analysis of the reflectance measurements collected from virgin and ocean-harvested plastics. Their findings showed that ocean-harvested plastics (ropes, foam, etc.) followed identical absorption features and had lower reflectance compared to virgin plastics (low-density polyethylene (HDPE, LDPE), polypropylene (PP)). Moshtaghi et. al.\cite{moshtaghi2021spectral} proposed one of the most important hyperspectral reflectance analyses of plastics in a controlled environment. The authors analysed reflectances of virgin and natural plastics submerged in water with different sediment conditions and depths. Their findings provided evidence to utilise shortwave infrared (SWIR) and visible spectrum for plastic detection.

Even though all the works published in this area can be somehow considered to be in computational imaging, in the remaining part of this chapter, we mostly focus on works that promote machine/deep learning-based computational imaging techniques such as feature extraction, segmentation, classification for the purpose of marine debris and floating plastics detection. 

\begin{table}[ht!]
    \centering
        \caption{Frequently Utilised Spectral Indices}
    \label{tab:indices}
    \begin{tabular}{p{1.5cm}p{6.5cm}p{9cm}}
    \toprule
        & \textbf{Index}		&\textbf{Expression}\\\toprule
\textbf{NDVI} &	Normalized Difference Vegetation Index&	 $\dfrac{R_{NIR} - R_{RED}}{R_{NIR} + R_{RED}}$\\\midrule
\textbf{FDI}& 	Floating Debris Index&	 $R_{NIR} - \left[R_{RE2} + (R_{SWIR1} - R_{RE2})\times \dfrac{\lambda_{NIR} - \lambda_{RED}}{\lambda_{SWIR1} - \lambda_{RED}}\times 10\right]$\\\midrule
\textbf{PI}& 	Plastic Index&	 $\dfrac{R_{NIR}}{R_{NIR} + R_{RED}}$\\\midrule
\textbf{WRI}& 	Water Ratio Index&	 $\dfrac{R_{GREEN} + R_{RED}}{R_{NIR} + R_{SWIR2}}$\\\midrule
\textbf{RNDVI}& 	Reversed Normalized Difference Vegetation Index&	 $\dfrac{R_{RED} - R_{NIR}}{R_{RED} + R_{NIR}}$\\\midrule
\textbf{AWEI} &	Automated Water Extraction Index&	 $4 \times (R_{GREEN} - R_{SWIR2}) - (0.25 \times R_{NIR} + 2.75 \times R_{SWIR1})$\\\midrule
\textbf{MNDWI}& 	Modified Normalization Difference Water Index&	 $\dfrac{R_{GREEN} - R_{SWIR2})}{R_{RED} + R_{SWIR2}}$\\\midrule
\textbf{NDMI}& 	Normalization Difference Moisture Index&	 $\dfrac{R_{NIR} - R_{SWIR}}{R_{NIR} + R_{SWIR}}$\\\midrule
\textbf{NDWI} &	Normalized Difference Water Index&	 $\dfrac{R_{GREEN} - R_{NIR}}{R_{GREEN} + R_{NIR}}$\\\midrule
\textbf{SAVI}&	Soil adjusted vegetation index&	$(1 + L) \times \dfrac{R_{NIR} - R_{RED}}{R_{NIR} + R_{RED} + L}$\\\midrule
\textbf{NDWI2}&	The second normalized difference water index&	 $\dfrac{R_{GREEN} - R_{RED}}{R_{GREEN} + R_{RED}}$\\\midrule
\textbf{NDBI}&	Normalised Difference Build-Up Index&	$\dfrac{R_{SWIR} - R_{NIR}}{R_{SWIR} + R_{NIR}}$\\\midrule
\textbf{FAI}&	Floating Algae Index&	 $R_{NIR} - \left[R_{RED} + (R_{SWIR1} - R_{RED})\times \dfrac{\lambda_{NIR} - \lambda_{RED}}{\lambda_{SWIR1} - \lambda_{RED}}\right]$\\\midrule
\textbf{S2}-\textbf{based}&	Sentinel 2 based index&	$\dfrac{ (R_{RE3} - R_{NIR}) + (R_{NNIR} - R_{NIR})}{R_{RE3} + R_{NNIR}}$\\
\bottomrule
    \end{tabular}
\end{table}

Acuna-Ruz et. al. \cite{acuna2018anthropogenic} propose an approach with the objective to improve the identification of macroplastics through remote sensing over beaches. The authors promote the utilisation of 8-band very-high resolution WorldView-3 satellite products. Five important ML approaches of Random Forests (RF), Support Vector Machines (SVM) with three kernel functions of linear, polynomial, and radial basis function, and Linear Discriminant Analysis (LDA) have been tested in order to classify the beach debris. The results show that the SVM approaches were mostly the best with detection accuracy around 85-90\%. 
Fallati et. al. \cite{fallati2019anthropogenic} propose an ad-hoc methodology for beach litter analysis based on the combined use of a UAV and the deep-learning software of PlasticFinder. The utilised software PlasticFinder is a commercial software of DeepTrace Technologies. It consists of several Convolutional Neural Networks (CNN) which are constructed to detect and quantify Anthropogenic Marine Debris (AMD). The approach in \cite{fallati2019anthropogenic} reaches a sensitivity value of 67\% whilst the positive predictive value of 94\%. Despite the considerable performance of the PlasticFinder, this software relies on the sunlight conditions of the scene and has serious problems with targets such as footprints or shadows.

The first satellite-only computational imaging approach has been proposed by Biermann et al. \cite{biermann2020finding} and they have utilised a Näive Bayes classification approach for the floating plastic classification. In fact, this approach has two main steps: (1) manual detection of marine debris using FDI and NDVI values, (2) using Näive Bayes to classify plastics among water, seaweed, timber, and foam. Across all five test sites, the proposed approach reaches 86\% detection accuracy of plastics of the pixels detected as marine debris in the first manual step. Kako et. al. \cite{KAKO2020111127} propose a novel approach which is a combination of UAV surveys and deep learning approaches. Via performing edge detection they estimate plastic marine debris on the beaches of Sato and Fukiage in Kagoshima, Japan. The computational imaging approach in \cite{KAKO2020111127} exploits a binary classification on each pixel (either plastic or not) based on 3 layers deep artificial neural network with a performance of less than 5\% error. 

Furthermore, Jakovljevic et. al. \cite{jakovljevic2020deep} aim to investigate the applicability of various important deep learning algorithms for automatic floating plastic via utilising UAV orthophotos. They also test to discriminate plastic types by exploring various characteristics such as spatial resolution and detectable plastic size. Utilised deep neural network approaches are semantic segmentation algorithms based on U-Net architecture. Results show that the ResUNet50 architecture achieved the best performance via detecting plastics with more than 85\% of precision.
An automated plastic pollution monitoring approach for the river surfaces using bridge‐mounted camera imagery is presented by van Lieshout et. al. in \cite{van2020automated}. The authors have performed an experimental analysis of five different rivers in Jakarta, Indonesia. The proposed deep learning-based approach consists of two cascaded CNNs for the purposes of segmentation and object detection. For the segmentation stage, the so‐called Faster R‐CNN (trained on the river image data set) was used whilst for the detection stage Inception v2 network (pre-trained on the COCO data set) was used. With the highest 69\% precision of plastic detection, their results demonstrate a promising way of monitoring plastic pollution.

Freitas et. al. \cite{freitas2021remote} propose a hyperspectral imaging-based ML procedure to detect floating plastics. Data set collected with manned/unmanned airborne hyper-spectral sensors have been tested with two supervised methods of RF and SVM. Their results show an automated detection capability of marine litter with up to 80\% precision (50\% recall).
Tasseron et. al. \cite{tasseron2021advancing} present a hyper-spectral laboratory setup to collect spectral signatures of 40 virgin macroplastic items and vegetation.  The system covers a spectral range from visible to shortwave infrared (VIS-SWIR) from 400 to 1700 nm. The experiment setup has given around 2 million pixels each of which has been processed via linear discriminant analyses (LDA). Their results return absorption peaks of plastics (1215 nm, 1410 nm) and vegetation (710 nm, 1450 nm), and provide evidence of the gold spectral indices of NDVI and FDI. 

Basu et. al. \cite{basu2021development} utilise Sentinel-2 multi-spectral imagery to detect floating plastic debris in coastal water-bodies in specific test locations in Limassol, Cyprus, and Mytilene, Greece. The utilised Sentinel-2 images are of the date on which the real-world plastic pollution data were collected through experiments at the two study area locations. The authors test four machine learning classification approaches K-means, fuzzy c-means (FCM), support vector regression (SVR), and semi-supervised fuzzy c-means (SFCM). In addition to the 6 Sentinel-2 bands of blue, green, red, red edge 2, near-infrared, and short wave infrared 1, two indices important spectral indices of NDVI and FDI are utilised to develop the ML models. Their results suggest that SVR outperforms all other ML models with around 97\% of plastic detection accuracy.

Kremezi et. al. \cite{kremezi2021pansharpening} leverage the potential of satellite hyper-spectra remote sensing imagery in marine plastic litter detection purposes for the first time in the literature. The authors use PRISMA satellite data with fine spectra but low spatial resolutions. In order to increase the spatial resolution, \cite{kremezi2021pansharpening} proposes exploiting pansharpening with the panchromatic data that enhance spatial resolution. In the experimental analysis, thirteen different pansharpening methods have been tested and the PCA-based substitution method managed to efficiently discriminate plastic targets from water-bodies. 

\begin{figure}[htbp]
    \centering
    \subfigure[]{\includegraphics[width=\linewidth]{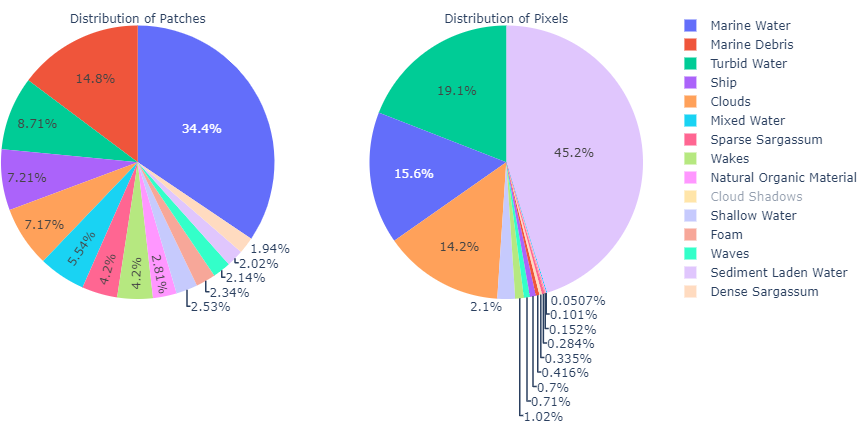}}
    \subfigure[]{\includegraphics[width=\linewidth]{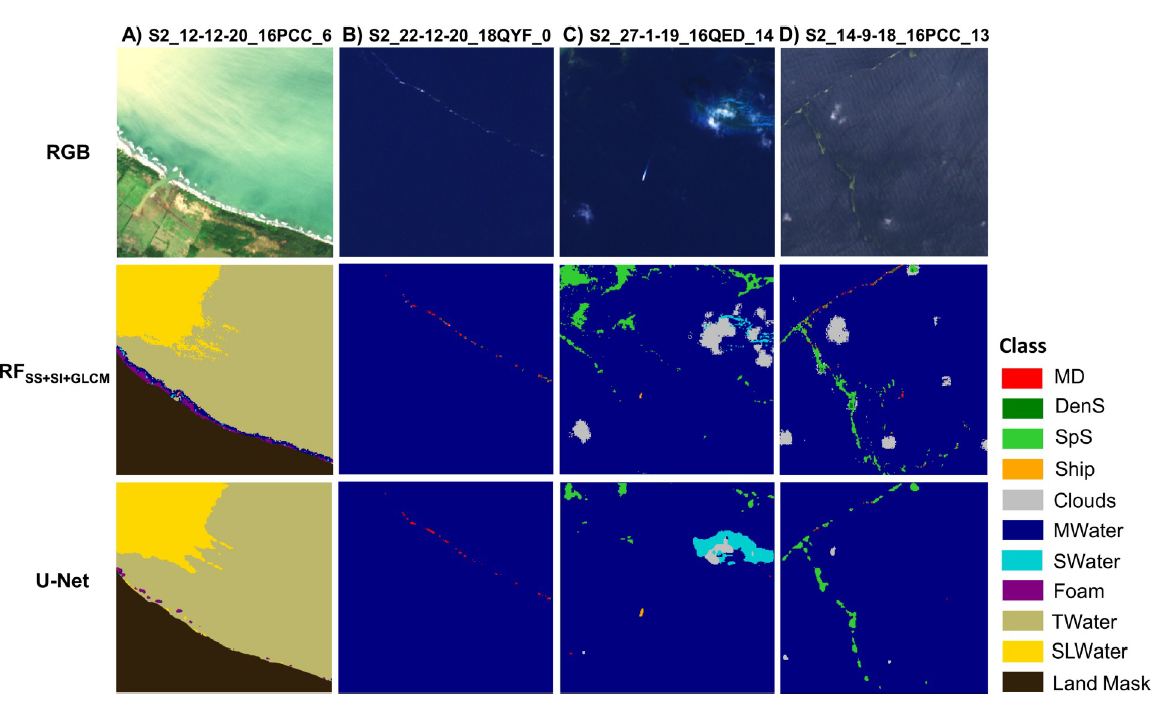}}
    \caption{MARIDA data set content statistics and example results for the benchmarking. (a) MARIDA data set class distributions. (b) An example performance result for the MARIDA benchmark models}
    \label{fig:marida}
\end{figure}

A very recent study by Kikaki et. al. \cite{kikaki2022marida} introduce a benchmark Sentinel-2 dataset for developing and evaluating ML algorithms, called Marine Debris Archive (MARIDA). Publishing the MARIDA data set will surely become a game-changer in marine debris and floating plastic detection research since it is the first dataset based on the multispectral Sentinel-2 satellite data. MARIDA offers the capability to classify various different marine targets, such as marine debris (MD), sargassum macroalgae, ships, wakes, foam, various dissimilar water types, and clouds. The authors have gathered plastic pollution information for the time interval between 2015-2021 from more than 11 different countries all around the world. Based on these ground truth data, Sentinel-2 images collected and annotations have been done via using high-resolution information from Planet and Google Earth imagery. MARIDA data consists of 17/63 Sentinel-2 tiles/scenes, 2597 patches, and 837377 annotated pixels. Among those pixels, MARIDA contains 3339 Marine Debris (MD) pixels in total. In the MARIDA paper, MD was defined as "floating plastic and polymers, mixed anthropogenic debris". Of these pixels, 1625 pixels were digitized and annotated with high confidence, 1235 pixels were labeled with moderate, and 539 pixels with low confidence. Confidence levels were determined by the collection of reports about mass-plastic events in coastal regions. All satellite images were also investigated with very high-resolution satellite imagery where possible. Overall, this increases the validity and makes MARIDA the most comprehensive classification of marine debris to date. From the technical point of view, MARIDA \cite{kikaki2022marida} also presents two different baseline approaches based on RF (has three versions for (1) spectra, (2) spectra \& indices, and (3) spectra \& indices \& land cover map) and U-Net architecture. The results indicate that RF forest approaches outperform U-Net results with $F_1$ scores higher than 0.7 whilst U-Net achieves 0.5. MARIDA data class distributions and an example performance result for the benchmark models are depicted in Figure \ref{fig:marida}. 

Kremezi et. al. in their recently published paper \cite{kremezi2022increasing} propose to enhance the capabilities of Sentinel-2 multi-spectral imagery via performing image fusion approaches with very high-resolution WorldView-2/3 (WV-2/3) images. Their approach has the primary aim to present solutions to the main limitation in the existing approaches that limited spatial resolution of Sentinel-2 hindering the detection of relatively small marine debris and floating plastics. Various image fusion techniques have been tested in \cite{kremezi2022increasing} in terms of preserving spectral and spatial information. Among those, Coupled non-negative matrix factorization (CNMF) has shown to be the best via producing a fused image with clear edges, no blurring, and favorable spectral characteristics. Utilised Fusion-GAN and Fusion-ResNet approaches have also shown great performance for fusion purposes regarding spectral similarity. The superior performance throughout the paper can be listed as the reduction in the required pixel coverage. The smallest detectable target via the best approach in the fused image was noted as 0.6 × 0.6 m2 in size, which is equivalent to 3\% pixel coverage of the original Sentinel-2 imagery with 20m resolution. Despite this notable impact and potential future research direction, their plastic detection approach consists of a high number of false detections. This is actually not special to this specific paper, but nearly all the existing approaches are suffering from false detections which reduces the accuracy and reliability of the detection approach. Considering proposing a detection approach is not the aim of \cite{kremezi2022increasing}, this paper can be noted as a breakthrough for this research area for future works. 

\begin{figure}[ht]
    \centering
    \subfigure[]{\includegraphics[width=0.8\linewidth]{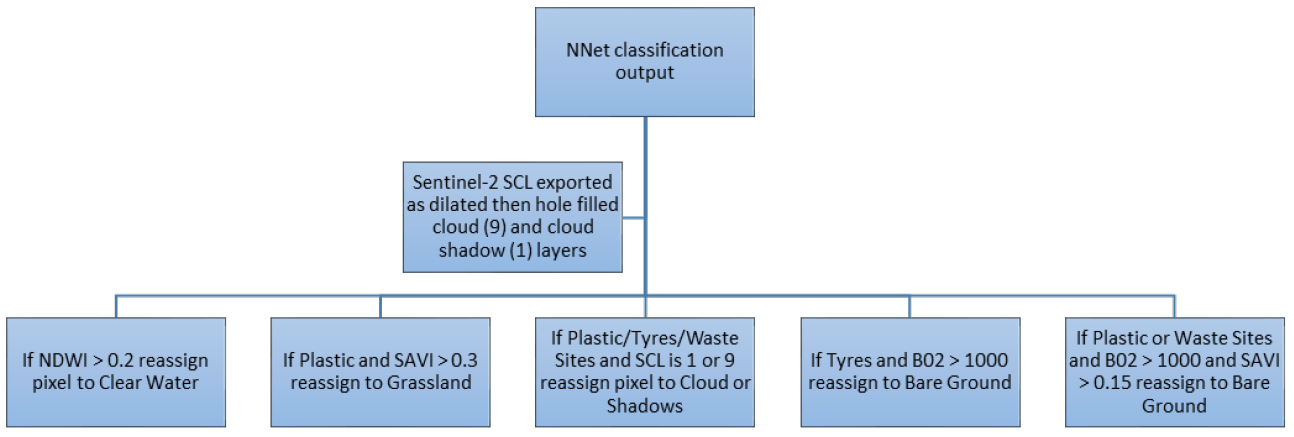}}
    \subfigure[]{\includegraphics[width=\linewidth]{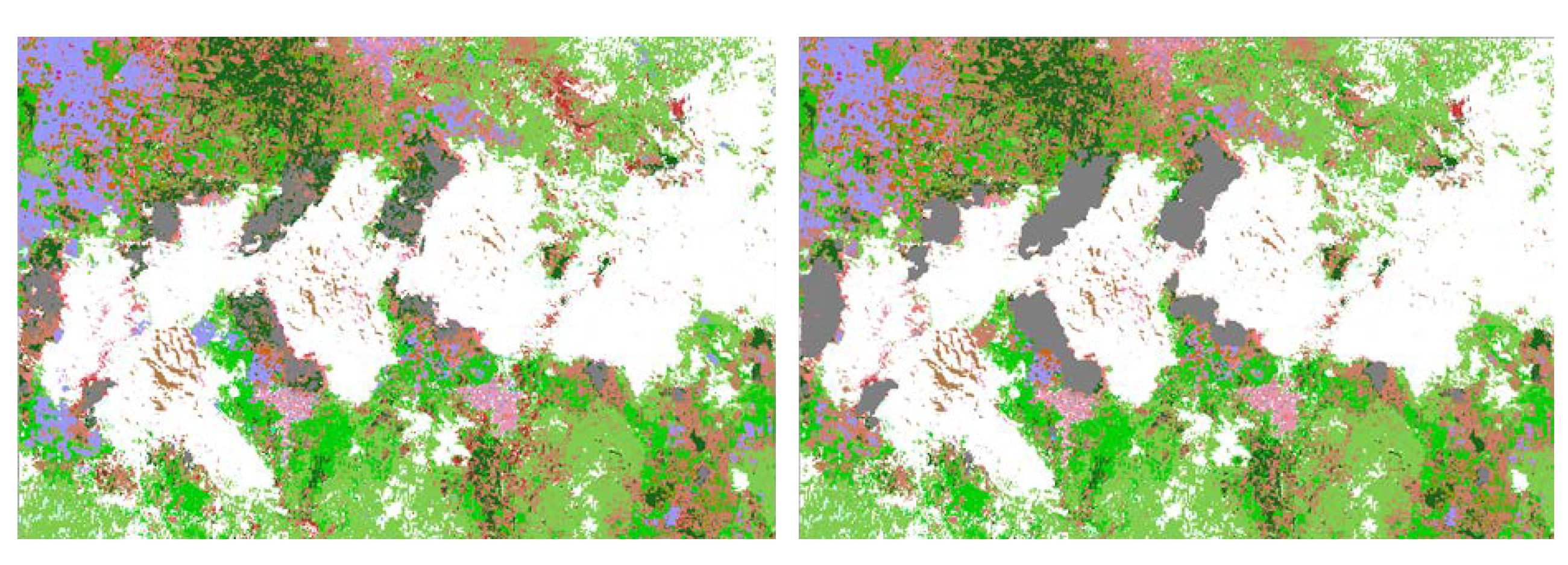}}
    \caption{(a) Post-ANN decision tree structure of \cite{lavender2022detection}. (b) Comparison of the classification results before and after the post-network decision tree for the subset of the Solo River.}
    \label{fig:kremezi}
\end{figure}

Lavender \cite{lavender2022detection} proposes a detection approach using satellite imagery, but this time not only for marine, also for terrestrial environments. This work also produces a data set by manually digitising some of the classes in the land cover maps into plastics, greenhouses, tyres, and waste sites. This data consists of around 1.3 million pixels only around 1\% of which is labelled as plastics. The proposed detection algorithm is fed by two different modalities of SAR (Sentinel-1) and multi-spectral (Sentinel-2) imagery and consists of an artificial neural network (ANN). An RF-based algorithm has also been used to compare the ANN performance. Except for the proposed data set, maybe the most important contribution of this paper in terms of the computational image processing approaches is to propose a post-ANN decision tree. This adds a second classification step to the proposed method in order to fix errors caused by the ANN's incapability to fully capture the radiometric interpretation of the surfaces and to process categorical layers. The post-ANN decision tree and an example before and after this has been shown in Figure \ref{fig:kremezi}. Regarding the detection performance, despite having around 95\% of aggregate average precision, their plastic detection precision is 91\%. 

\begin{figure}[t]
    \centering
\subfigure[Manila, Phillippines]{\includegraphics[width=.49\linewidth]{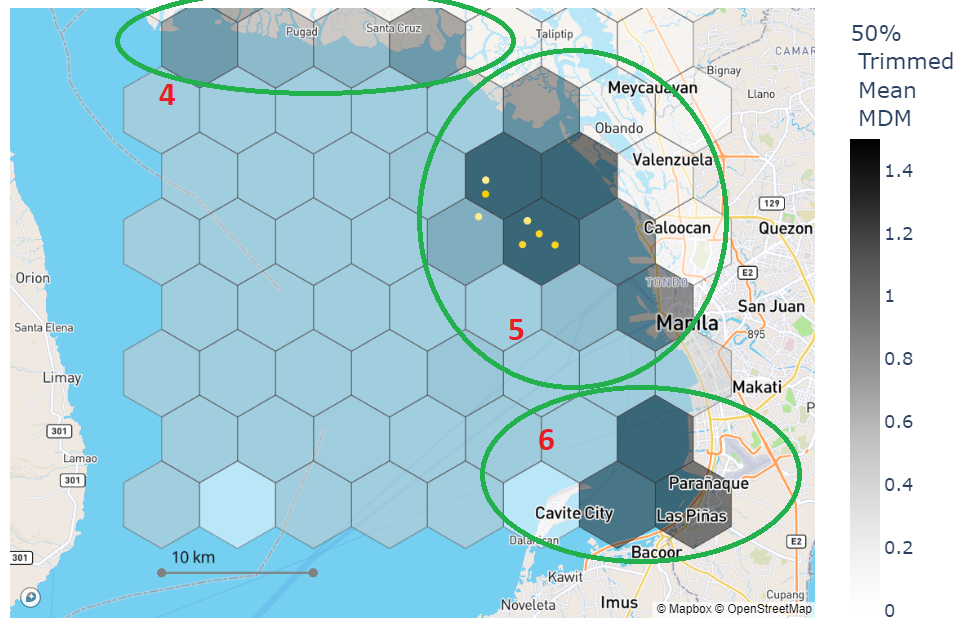}\label{fig:manila}}
\subfigure[Mumbai, India]{\includegraphics[width=.49\linewidth]{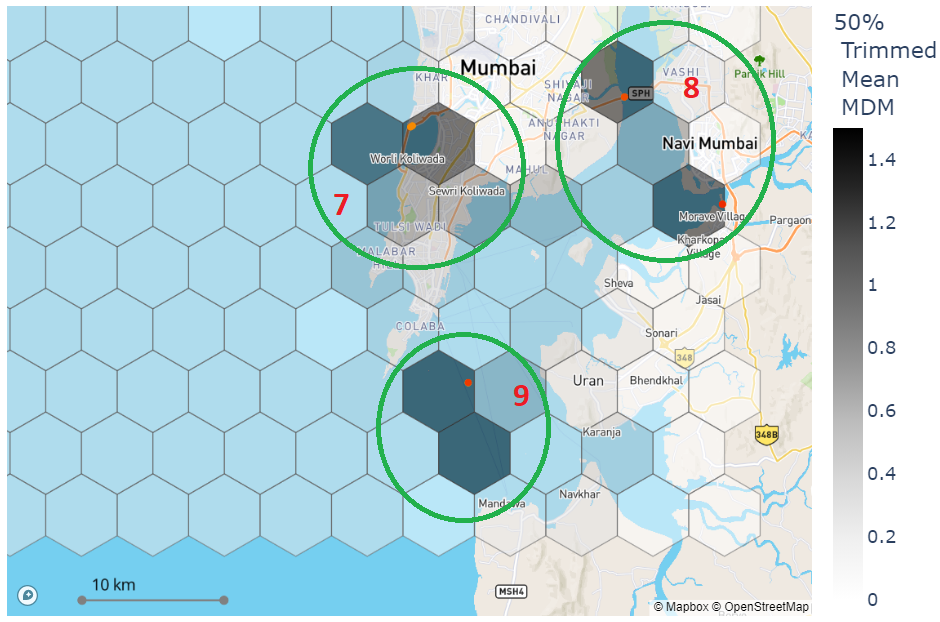}\label{fig:mumbai}}
    \caption{Marine Debris Density Map from \cite{booth2022high} for (a) Manila, Phillippines, and (b) Mumbai, India. The density maps show results for a one-year time interval in 2022. The colour bar corresponds to 50\% left-trimmed average MDM values of each hexagon. Each figure also includes the top 10 highest pixel-level MDM values as scatter points where brighter markers correspond to higher MDM values. }
    \label{fig:henry}
\end{figure}

Booth et. al. in their work \cite{booth2022high} propose a high-precision marine debris mapping algorithm by utilising Sentinel-2 imagery and a semantic segmentation algorithm. Their approach is constructed to detect marine debris and floating plastics from a high-precision point of view. Maybe, in the literature, it is the first time, this approach focuses on improving the precision of the marine debris detection algorithms by keeping recall values high enough to produce a reliable algorithm. The authors in \cite{booth2022high} facilitate semi-automated monitoring of suspected marine debris and floating plastics via a data pipeline named MAP-Mapper that enables the user to provide coordinates for a region of the ocean along with a date time range so that Sentinel-2 data could be downloaded, pre-processed via atmospheric corrections and individual pixels could then be classified using the proposed ML algorithm. Afterwards, as of the outcome of this tool, a plastic density map is produced and enables users to easily interpret model predictions and identify areas of high plastic density in the investigated area. The authors also proposed a novel index called the Marine Debris Metric (MDM) which is a positively valued metric where 0 means no marine debris problem, and a higher $MDM$ value corresponds to a polluted location on the maps. On the other hand, $MDM$ can be seen as a metric in which the average probability value of a pixel is positively weighted if the total number of detections is high in that corresponding pixel (e.g. 10 dates out of 20 dates in the time interval means 50\%). This gives us a better picture of measuring the MD\&SP density maps in a global time scale with the MAP-Mapper approach. $MDM$ can also be seen as a universal metric to compare different locations on the earth considering the utilisation of the fixed area of hexagons and the averaging over the given time interval. Their approach promotes utilising the MARIDA data within a U-Net-inspired machine learning approach. Their system MAP-Mapper-HP reached 95\% precision of marine debris class in MARIDA data set whilst the original U-net based baseline in \cite{kikaki2022marida} was only 30\%. In order to test their approach's applicability, they set six out-of-distribution test locations such as the Bay of Honduras, Cornwall, and Mumbai - India via the data pipeline. Their generated marine debris maps for these test locations are aligned with the existing marine debris gathering information and ground truths. In Figure \ref{fig:henry}, example marine density maps for Mumbai and Manila are depicted.

Apart from the hyper- and multi-spectral remote sensing based approaches mentioned above, the literature also accommodates various recently published active sensor data promoting research on marine plastic detection. In the work published in 2021 by Savastano et. al. \cite{savastano2021first}, one of the first successful applications of SAR imagery to marine debris detection problems is proposed. The authors developed an AI system to utilise Sentinel 1 SAR data as input along with Sentinel 2 multi-spectral data as ground truth. Their test location is the Balearic islands, an archipelago located within the Western Mediterranean Sea. This area is selected due to the reason that it currently is a hotspot area for plastic marine debris accumulation situated between the two dominant currents, the Algerian and the Balearic currents. They promote utilising Gaussian Naive Bayes, RF and SVM approach where they reach up to 86\% accuracy. In another work by Serafino and Bianco \cite{serafino2021use}, a ground-located X-band radar sensor has been used to identify, discriminate, characterise and follow small floating aggregations of marine litter made up mainly of plastic debris. Their results provide valuable conclusions for the future usage of radar sensors in marine debris research. They concluded that in calm sea conditions (almost no wind), X-band radar was capable of distinguishing experimental targets on the sea surface within the range of 0.39 nautical miles. In a very recent research outcome, Giusti et. al. \cite{giusti2022drone} proposed a drone-based multi-sensor system for marine debris detection. The authors present the main results of the POSEIDON project in which a commercial drone has been equipped with a radar and a multispectral camera jointly observing a common area of interest. This approach promotes utilising an image fusion approach to exploit the advantages of both radar and multi-spectral imagery. The proposed detection algorithm for multispectral imagery leverages the use of a CNN with a multi-resolution Feature Pyramid Network (FPN) backbone. For SAr data, a constant false alarm (CFAR) based detection mechanism is proposed.

\section{Literature Statistics}\label{sec:stats}
This Chapter is going to highlight the evolution of marine debris research with numbers in terms of literature statistics. Our analysis will follow a gradual point of view - from general to special - in three cases that titles, abstracts, and/or keywords include the words: 
\begin{enumerate}
    \item ("Marine Debris"),
    \item  ("Marine Debris") AND ("Remote Sensing" OR "Satellite" OR "Sentinel"),
    \item ("Marine Debris" OR "Plastic") AND ("Remote Sensing" OR "Satellite" OR "Sentinel") AND ("Machine Learning" OR "Deep Learning" OR "Artificial Intelligence")
\end{enumerate}

Please note that the keyword choice above is assumed to be one of the most feasible ones for the purposes of this review. Other kinds of choices might create different statistics but this case is out of the scope of this review paper.

All the statistical analyses are performed via the document search option in Scopus (\url{www.scopus.com}) and the presented data in this section is collected on the 15$^{th}$ February 2023. This means literature pieces that are not indexed via Scopus are out of the scope of this analysis. 

\subsection{Marine Debris Research}\label{sec:case1}
\begin{figure}[ht!]
    \centering
    \subfigure[Country based percentages]{\includegraphics[width=0.69\linewidth]{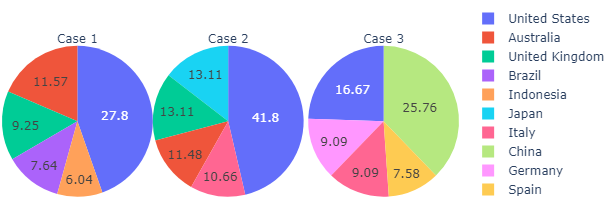}}
    \subfigure[Yearly trend]{\includegraphics[width=0.49\linewidth]{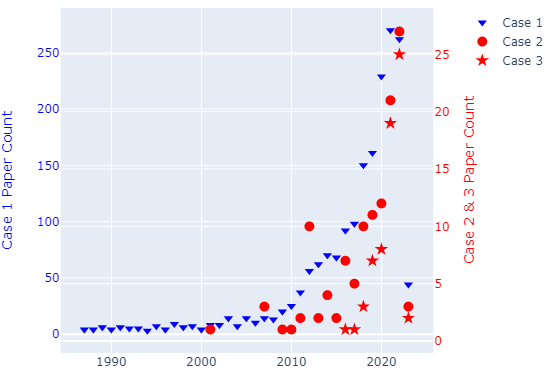}}
    \subfigure[Subject area percentages]{\includegraphics[width=0.49\linewidth]{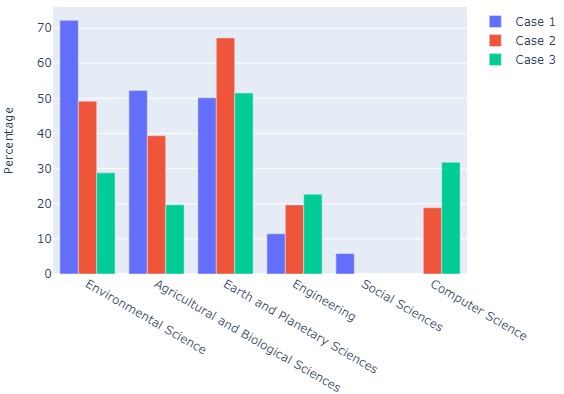}}
    \subfigure[Source title based percentages]{\includegraphics[width=0.9\linewidth]{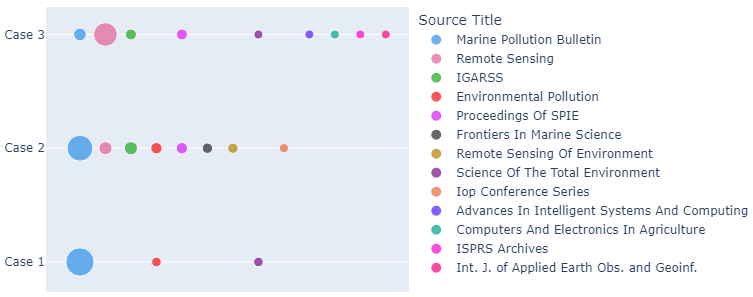}}
    \caption{Data visualisation for the literature statistical analysis.}
    \label{fig:cases}
\end{figure}

Marine debris research dates back to the 80s according to our analysis which can be seen in Figure \ref{fig:cases}-(a). Even though the number of papers is quite low, it can be taken as evidence of how old marine debris problems and attracted attention in academia. This slow trend continues still for nearly 30 years until the 2010s. 2010 is a critical point since 90\% of the papers published after 2010 in around 13 years. 2019+ is another critical year that more than 50\% of the works have been published during the last 5 years of time. The main important reason for this is the improved technology in both remote sensing satellite products and artificial intelligence research.

In terms of the subject area, three important areas appear to lead this research, that is Environmental Science, Agricultural and Biological Sciences, and Earth and Planetary Sciences. United States of America (USA) appears to be the leading country in this research via populating approximately 28\% of the papers whilst Marine Pollution Bulletin becomes the leading academic source title via accommodating more than 30\% of the published papers. 

\subsection{Marine Debris Research via Remote Sensing}\label{sec:case2}
When we tighten our search keys by adding remote sensing and some other related terms, Scopus returns 122 academic outcomes. Compared to Case 1, there is a dramatic decrease in the number of papers. The main reason is we now specifically look at marine debris research that promotes utilisation of remotely sensed data, such as RGB, optical, and radar imagery. Along with the remote sensing research, Case 1 also includes various important marine debris research in areas of biology, material sciences, chemistry, etc. Now, most of these research are excluded unless they are promoted using remote sensing imagery. It looks like Scopus only indexed 5 works within these keywords before 2010, which makes it again an important milestone date for marine debris research via remote sensing imagery. Similarly, around 60\% of the papers that are highlighted in this case were published in 2019 and thereafter. 

The three important subject areas which become prominent for Case 1 are still the best three subjects. However, this time, adding remote sensing details into the play made Earth and Planetary Sciences research area the leader naturally. The USA is still the leading country in this research with more than 41\% papers published by US institutions. Similarly, the Marine Pollution Bulletin ($\approx$26\%) journal is a far more widespread academic venue high-likely to see marine debris research via remote sensing imagery. 

\subsection{Marine Debris Research via Remote Sensing Computational Imaging}\label{sec:case3}
Case 3 aims to highlight research at the intersection of oceanology, earth sciences, and computer science. We can also mention that the research highlighted under Case 3 is the target research of this review paper. Adding deep/machine learning keywords into the search criteria basically returns us to remote sensing computational imaging approaches that aim to provide solutions to marine debris and floating plastics problems. Our Scopus search returns 66 academic outcomes for Case 3 where the first of which was published in 2016. Computer vision research and the AI revolution can be dominantly felt in the marine debris research area, especially during the last 3 years in which more than 80\% of the works were published.

Earth and Planetary Sciences research area is still the leading area whilst Computer science and Environmental Sciences follow it with a high amount of work. There are two important breaking points different from the statistical analyses for Cases 1 and 2. In this case of research, China has become the new leading country whereas USA institutions still publish around 17\% of the papers. It is also important to note that European institutions in Italy, Germany, Spain, and Greece have started to direct this research area by publishing a total of 30\% papers. 
Taking AI into play made The MDPI Remote Sensing journal (a peer-reviewed, open-access journal about the science and application of remote sensing technology) the leading academic venue for this research whilst Marine Pollution Bulletin and the International Geoscience And Remote Sensing Symposium (IGARSS) follows with a similar number of publications. 

\subsection{Fraction of Marine Debris Studies in Remote Sensing Research}
Finally, despite all the efforts statistically mentioned above, marine debris research via remote sensing computational imaging is still in the early stages. In order to understand this, we share two important Scopus searches and the numbers will say it all!

When we remove ("Marine Debris" OR "Plastic") keys from Case 3, Scopus returns 26393 academic outcomes. This clearly means that the marine debris research that promotes remote sensing computational imaging approaches is only 0.25\% within the whole research area. When we go one step down by excluding some non-marine related research from this analysis and replacing ("Marine Debris" OR "Plastic") key in Case 3 with ("Marine" OR "Maritime" OR "Ocean" OR "Water"), Scopus returns 4144 academic outcome leads only to 1.59\% of the research in this area promotes remote sensing and computer vision approaches for marine debris monitoring purposes. 

It is clear from the above discussion that most of the studies (and therefore, users and customers) are not directly targeting marine debris detection. This is one of the natural reasons for the fact that the existing satellite sensors and missions have not been designed to monitor, detect, and track marine debris which appears as a great necessity to change in the future!

\section{Challenges} \label{sec:chal}
This chapter draws a picture regarding the potential challenges in marine debris and floating plastics detection research area from the point of view of the authors. We are going to reflect on our understanding and experience within this chapter and please note that the challenges reported below are not exhaustive.

\subsection{Optical Imagery Limitations}
As we explained in the previous literature review that the works highlighted mostly focus on optical multi-hyper spectral imagery, which provides a great amount of information whilst having some serious disadvantages at the same time. There are several reasons that can be listed as potential challenges for passive remote sensing systems in marine plastic pollution monitoring.
\begin{enumerate}
    \item As mentioned in the above discussion, even though providing data with 13 different spectral bands, the current gold modality Sentinel-2 imagery has up to 10 meters of spatial resolution. This significantly influences the detection capability of methods since the amount of plastic in a single pixel determines the light intensity and defines a lower size limit for floating plastics, which are around 30-50 m2. This lower limit of plastic is quite high considering that the ocean generally has various pollutants combined in a single patch. The sub-pixel indexes (the FDI of \cite{biermann2020finding}) can be thought of as a solution to this problem, but it is still open to improvements for higher performance. Image fusion approaches that exploit high-resolution data along with the Sentinel-2 information will be important future research direction as Kremezi et. al. \cite{kremezi2022increasing} have given an example for this via utilising a similar approach which leads to a reduced required pixel coverage of 0.6 m by 0.6 m.  
    \item In addition to the previously mentioned fact that optical images are prone to cloud cover and incapable of night-time data generation, optical remote sensing also suffers from lower sampling rates (say high revisiting times). All these reasons limit their capability to collect continuous data, specifically for plastic patches tracking applications. To the best of our knowledge, in the literature, there is a limited number of works that have studied the problem of tracking plastic patches in the oceans. There are several but non-exhaustive reasons behind this, which are required to be addressed. 
    \begin{enumerate}
        \item In addition to the incapability to obtain continuous optical imagery for tracking purposes, due to atmospheric limitations of the optical sensors, the low spatial resolution of optical sensors also causes a risk in tracking individuals of plastic bits. 
        \item Considering the floating plastics are drifting due to winds, sea waves, and currents, developing a robust tracking approach requires modeling these hydrological variables. Since optical sensor wavelengths are not capable of covering the Bragg scattering mechanism, imaging hydrological variables such as gravity waves, swell waves, and ocean currents cannot be possible.
    \end{enumerate}
    \item The limitations of the optical satellite remote sensing data have led the plastic detection research somehow to unmanned air vehicle (UAV) based approaches. Although these works suggest important outcomes for the understanding of the remote sensing imaging of plastic patches \cite{moy2018mapping,garaba2018airborne,topouzelis2019detection}, they are operator and location dependant whilst having high costs and lack of standardisation. 
\end{enumerate}	

\subsection{Data Related Problems}
In addition to the sensor-related challenges mentioned above, there is also another important problem causing this research area to move slower compared to other computational imaging areas; that is \textit{the data availability and other data-related problems}. 

The very first one is the limited number of open-access marine debris and plastic detection data sets. To the best of our knowledge, before 2022, there was only one available data set from Topouzelis et. al. \cite{topouzelis2020remote} under several repetitions of plastic detection contests. Despite providing a great source of information for future studies as is mentioned in the above chapters, the shared data set consists of a couple of Sentinel-2 imagery that includes only a handful of annotated plastic pixels. Thus, its applicability in advanced remote sensing computational imaging approaches naturally remained limited. Thanks to the efforts in this area, during 2022 (very recently) two important data sets have been published \cite{kikaki2022marida,lavender2022detection}. These data sets, even though still suffering from the number of annotated plastic pixels, provide a basis for researchers around the world and we believe that the applicability of the advanced deep-machine learning approaches will be dramatically increased thanks to these two data sets. 

As of now, despite having several data sets, we still face an important problem, which is the limitation of annotated plastic and marine debris pixels. Due to their locations in the ocean, and accessibility issues, having ground truth information with a high level of confidence is a great challenge for this research area. The MARIDA data set \cite{kikaki2022marida}, despite having around 0.9 million annotated pixels, only 0.4\% of those belong to the marine debris class. Among those, only half of these pixels (1625 pixels reported in \cite{kikaki2022marida}) are annotated as high confidence which makes the reliable information in the data set even smaller. This limited information causes a lack of learning for the advanced algorithms and we believe this is the reason that nearly all the proposed approaches suffer from low precision, in other words, lots of false plastic/marine debris detections. 

Lastly, both having a limited number of data sets and a small amount of high-confidence annotated pixels, advanced computer vision algorithms have not been utilised for the purpose of detecting floating plastics and marine debris. Instead, most of the referenced works above rely on classical approaches such as Random Forests, Naive Bayes, or simple machine learning architectures based CNNs. 

Interested readers may refer to the exhaustive list of databases which is created by the IOCCG\footnotemark[15] covers remote sensing and marine debris/litter keywords.
\footnotetext[15]{\url{https://ioccg.org/rsmld-datasets-bibliography/}}

\section{Future Research Directions} \label{sec:future}
After highlighting several important challenges in this area in the above section, now, we share potential future research directions.

\subsection{Utilisation of SAR Imagery}
Considering the limitations and challenges mentioned above, it should also be admitted that the detection and monitoring of plastic pollution via remote sensing imagery are still in their early stages, especially in terms of multi-modality and promoting other sensor information. All the efforts in the last decade, lead to the development of the pioneering works mentioned above. Optical imagery, though being the modality that displays the earth as the human eye sees it, will, unfortunately, be at the mercy of the clouds and the amount of sunshine. However, the SAR system constitutes an active sensor, which emits microwaves towards the Earth and receives back-scattered signals from the surface. Since SAR utilises larger wavelengths (1cm to 1m) compared to the optical sensors (near-visible light or 1 micron), it is able to image both day and night, and in almost all-weather conditions through clouds and storms (while optical sensors are not). SAR imagery provides us with useful information about forest biomass, crop cover types and their ability to sensitively discriminate against the target surface. Furthermore, SAR images also provide useful information on ocean state, including wind speed and direction, gravity waves, swells, currents, sea-ice structures, and meteorological and environmental conditions in bodies of water. Thanks to its aforementioned capabilities, SAR imagery is recognised as the gold standard and superior to optical imagery where the literature abounds various applications which utilise SAR imagery as the main source of information, such as detecting/tracking oil spills \cite{brekke2005oil, karantzalos2008automatic, kostianoy2006operational, li2010oil, shirvany2012ship, zhao2019detecting}, assessment of sea surface signatures \cite{karakucs2020rici, karakucs2019ship2, karakucs2019sim, graziano2016wake}, ship detection and tracking \cite{gao2020autonomous, kartal2019ship, liang2019destination, mazzarella2013data, pappas}, and ice sheets tracking.

SAR imagery has various advantages compared to the passive remote sensing modalities, which can be listed below:
\begin{enumerate}
    \item Several European missions (e.g., the Italian COSMO/SkyMed, the German TerraSAR-X, the UK's NovaSAR-1, or the Finnish ICEYE) have developed a new generation of satellites exploiting SAR to provide spatial resolutions previously unavailable from space-borne remote sensing (up to 0.25 meters). Investigating cm-scale pixel sizes via remote sensing modalities is crucially important to detect and classify the plastic pollutant since these spatial resolutions are not suitable for optical remote sensing, and operating UAV-based approaches with cm-scale resolutions are extremely expensive, operator and location dependent. Conversely, high-resolution SAR data is continuously obtained for some other monitoring applications regardless of our intention to use it. Taking these high-potential, yet less-demanded, sources into play for plastic detection yields crucial importance for the future role of remote sensing technologies in plastic pollution monitoring.
    \item New SAR technology provides larger swath-ranges compared to optical imagery. This is one of the reasons that makes SAR preferable to optical imagery, specifically for monitoring applications.
    \item Thanks to its day-night imaging capability, SAR provides up to 12 hours of temporal resolution whilst the Sentinel-2 optical imagery has a revisiting time of only 2 to 5 days. High-sampling rates, high temporal resolution, or smaller revisiting intervals can be listed as the primary need for monitoring/tracking plastic patches in the ocean. Utilising this source of information would help us to develop robust tracking algorithms. On the other hand, this also will serve the community with the understanding regarding how the plastic patches behave in the open ocean, and why places such as the remote islands at the Pacific Ocean like Henderson Island becomes a gathering point for ocean plastics \cite{lavers2017exceptional}.
    \item SAR sensors operate with various frequency bands (P-S-L-C-X), and each of these bands has various advantages compared to the others. As we mentioned above, SAR has been analyzed in a limited number of works for plastic detection and recognition in the literature to date. However, these analyses are limited only to Sentinel-1 SAR imagery, which images C-band data with 10-m spatial resolution. Since (i) different frequency bands will therefore develop different interactions between the plastic pollutant and the sea surface waters, and (ii) different penetrable frequency bands such as S (NovaSAR-1), L (ALOS2) and X (TerraSAR-X, ICEYE, etc.) have never been utilised in these analyses, SAR imagery is yet unexplored regarding its plastic detection capability. An example demonstration of the penetration capability of different SAR frequency bands is shown in Figure 4. This demonstration shows the L-band penetration capability to e.g., determine forest floods. To date, a similar analysis has not been studied for sea surfaces including plastics and various other natural materials. 
\end{enumerate}
\begin{figure}[htbp]
    \centering
    \includegraphics[width=\linewidth]{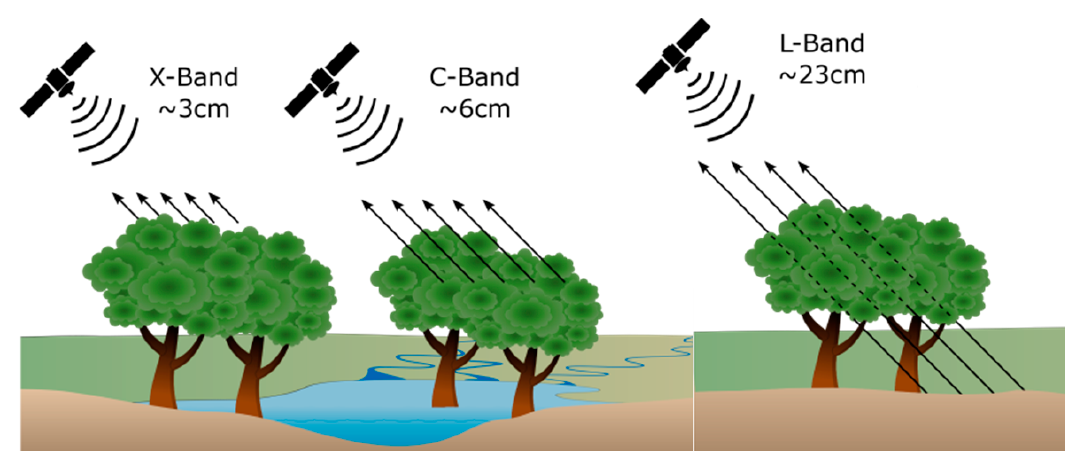}
    \caption{SAR frequency band penetration example \cite{ottinger2020spaceborne}.}
    \label{fig:sarFreqBands}

\end{figure}


Drawbacks: Despite the advantages compared to optical imagery, an important problem degrading statistical inference from SAR imagery is the presence of multiplicative speckle noise, which is not the case in optical imagery. The received back-scattered signals sum up coherently and then undergo nonlinear transformations. This in turn causes a granular look in the resulting images, which is referred to as speckle noise \cite{kuruoglu2004modeling}. Speckle noise may lead to the loss of crucial details in SAR images and cause problems in the processing of these images such as in feature detection, segmentation, or classification. However, speckle noise is a well-studied problem for decades and can be dealt with and removed by precisely determining the statistical characteristics of SAR images as in \cite{achim2006sar, karakucs2019cauchy2, karakucs2020GGRician, karakucs2018generalized, kuruoglu2004modeling}. On the other hand, even though SAR can provide very important information with added polarization and other capabilities, at the same time, surface waves present a major source of noise which is partly the reason why there are not many reports of successful debris detection.

\subsection{Applicability of New Generation Satellites}
\begin{figure}[htbp]
    \centering
    \includegraphics[width=\linewidth]{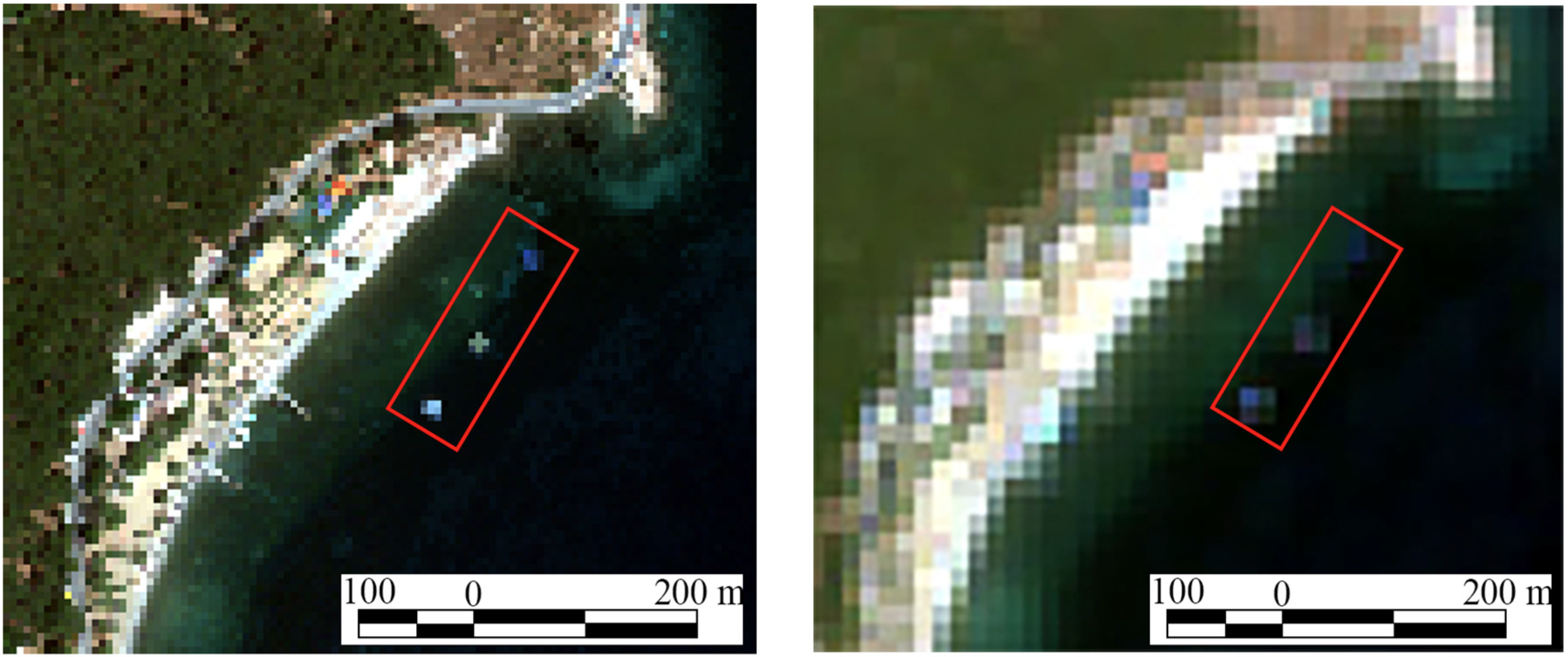}
    \caption{(left) RGB WV-3 image with 4 m spatial resolution, (right) S2 image with 10 m spatial resolution. \cite{kremezi2022increasing}.}
    \label{fig:wv3}
\end{figure}

Mentioning the drawbacks of optical multi/hyperspectral imagery above does not specifically mean that their era has already passed. Apart from the new technology satellites mentioned above for SAR imagery, in the last couple of years, new technology satellites for optical imagery have also been launched. Worldview-3 which is an imaging and environment-monitoring super-spectral, high-resolution commercial satellite sensor from Maxar, can be named among those technologies. Even though it has 8 years of history, due to being a commercial satellite, its usage has been limited to various remote sensing applications let alone marine debris monitoring. This instrument collects images at 0.31 meter panchromatic and 1.24 meter in the eight VNIR bands, 3.7m in the eight SWIR bands whilst having additional bands for enhanced multi-spectral analysis (coastal blue, yellow, red edge, NIR2) designed to improve segmentation and classification of land and aquatic features. Since Worldview-3 products started to be delivered by the ESA, their applicability has increased, and recently researchers reduced the smallest detectable marine debris target to 0.6 × 0.6 m2 in size, which is equivalent to 3\% pixel coverage of the original Sentinel-2 imagery with 20m resolution. We believe satellite products like Worldview-3 multi-spectral imagery will direct following years of research for floating plastics and marine debris monitoring via filling the gaps in low Spatio-temporal resolution of the current gold standard - the Sentinel-2 imagery products. An example of WV-3 and S-2 comparison for the same area is given in Figure \ref{fig:wv3}. 

\subsection{Potentials of Other Remote Sensing Sensors}
In addition to the potential applicability of SAR imagery for the purposes of marine debris and suspected plastic detection, there is some other sensory information which can be beneficial for the same purpose. Apart from the direct imaging satellites in the earth's orbit, there are also some other important satellites for measuring meteorological aspects of the earth from space. This information might potentially be used for the purpose of detecting and imaging the global distribution of microplastics. Evans and Ruf in their recent work \cite{evans2021toward} published in 2022, proposed a technique to detect and image microplastics with a spaceborne radar. Their method utilises a spaceborne bistatic radar that measures ocean surface roughness in order to estimate the reduction in responsiveness to wind-driven roughening caused by surfactants that act as tracers for microplastics near the surface. The efforts utilising meteorological radar systems to detect and map marine litter were potentially motivated by the seminal work of van Sebile et. al. \cite{van2015global} in 2015. The authors promote using a statistical framework for plastic marine debris measurements. They use surface-trawling plankton nets and coupled this with three different ocean circulation models to spatially interpolate the observations. The findings in this paper motivated the works in \cite{evans2021toward, van2020physical, van2022identifying}.

Different from other remote sensing techniques form raster imagery, the active remote sensing technique - light detection and ranging (LiDAR) - provides a vector dataset which is a cluster of points also named point-cloud datasets. LiDAR is basically a laser sensor the backscattered information of which is then digitised and discretised to form spatial 3D features of real objects. LiDAR data has relatively higher spatial resolution thanks to the gathered point cloud data, and until now it has been actively utilised in applications such as forestry \cite{dalponte2008fusion, wulder2013status}, land and coastal characterisation \cite{dubayah2000land, brock2009emerging}, biomass/biodiversity \cite{lefsky1999surface, zolkos2013meta}, atmosphere \cite{weitkamp2006lidar, bhardwaj2016lidar}, archaeology \cite{chase2012geospatial}. For the purposes of marine debris detection and recognition, LiDAR has also been utilised in the literature. In a seminal work by Ge et. al. \cite{ge2016semi}, a semiautomatic recognition of marine debris on a beach has been studied. Their results revealed that LIDAR is a useful tool for the classification of marine debris into plastic, paper, cloth and metal. In addition, the utilisation of LiDAR gave the capability to 3D model different types of debris on a beach with a high validity of debris revivification. Furthermore, in \cite{YANG2023103149}, Yang et. al. utilises terrestrial laser scanning to detect and extract marine litter in a coastal environment. 

All the aforementioned remote sensing techniques mentioned in this sub-section have the capacity to contribute to marine debris and suspected plastic detection research. Due to the small target sizes of the plastic pollutants on the sea surface, imaging technologies were generally suffering from the spatial resolution problem as is mentioned in the above sections. Remote sensing information apart from multi-, hyper-spectral and radar imagery has a great potential to develop improved performance approaches in the future. They might not be seen as useful sources individually, but would play an essential role especially in developing further multi-modal approaches that exploit all kinds of remote sensing information.

\subsection{Computational Imaging Solutions}
As we mentioned previously, this paper aims to review and discuss current advances in computational remote sensing imagery for the proposes of marine debris and floating plastics detection. In Section \ref{sec:pollution}, we specifically discussed computational imaging approaches utilised for this purpose, the conclusion leads to an important fact that existing approaches mostly rely on classical machine learning approaches such as Random Forests, Naive Bayes, and straightforward deep neural network structures. However, one of the most active research areas of deep learning and artificial intelligence is computer vision and relatively advanced approaches are being developed nearly every day. Despite this fast-changing and active nature of computer vision, remote sensing part of it still coming from several years back. Several reasons for this have been highlighted above as a lack of open access data sets, and a low amount of annotated plastic pixel existence. 

The first potential computational imaging research direction for marine debris monitoring that would be active in the following years is multi-modal AI and multi-sensor image fusion approaches. Various environmental applications have benefited from multi-modal image fusion by exploiting complementary features provided by different types of remote sensors. Specifically, for active-passive sensor data fusion, passive-optical sensors play the role of feeding the system with a high spectral resolution of the earth's surface which are useful for image analysis whilst active remote sensing sensors usually provide sufficient textural and structural information of observed objects \cite{zhang2022multispectral}. Some example remote sensing applications that exploit multi-sensor information and obtain considerable performance increase can be listed as land use/cover mapping \cite{ma2022amm}, air pollution detection \cite{scheibenreif2022toward}, building footprint extraction \cite{shermeyer2020spacenet}, and maritime vessel detection \cite{farahnakian2020deep}. 

Another computational imaging research direction will be guided by the fact that the marine debris monitoring area is suffering from a low amount of annotated data. The same existing fact has been a great driving force for the development of semi-, weakly-, and self-supervised learning (reduced supervision or minimal supervision) computational imaging approaches especially in the medical imaging area, which has the same annotated data problem as remote sensing imagery applications. Along with the advances in multi-modal fusion approaches, developing minimal supervision techniques for marine debris monitoring is a crucial step, and we believe academic literature will gradually get involved in this area in order to leverage not annotated data simultaneously with the limited amount of high-confidence marine debris pixels. The weakly supervised approach proposed in \cite{kikaki2022marida} is a good starting point for this, but of course not enough for high-precision marine debris monitoring software development tools. Ma et. al. in their land cover mapping application paper \cite{ma2022amm} have shown that without losing much performance accuracy, multi-modal sensor information usage in a parallel manner can reduce the amount of required training data up to 1/20. We believe this work, and other similar minimal supervision-promoted remote sensing computational imaging approaches in the literature might guide researchers in the following years. 

\subsection{Tracking the Source of Pollution}
To the best of our knowledge, the literature for marine plastic pollution monitoring does not yet have a complete tracking method due to the incapability of continuous data flow from optical sensors. On the other hand, resolution limitations and small sizes of plastic bits can be listed as other reasons for not developing a complete set of tracking approaches. The tracking literature is mostly dominated by the Bayesian approaches, especially for the last couple of decades that are based on the sequential Monte Carlo (SMC) algorithm. There have been proposed various advanced versions of this algorithm, though the SMC-based tracking approaches still suffer from data degeneracy and high particle variance problems. Currently, optimal transport theory-based solutions are getting prominent in tracking approaches \cite{corenflos2021differentiable,wu2022ensemble,hao4129781particle}. However, there is no such advanced methodology developed for remote sensing imagery so far. Thus, the marine debris tracking literature has an open and hot topic via the improved and novel version of the sequential Monte Carlo (SMC) algorithm, whilst leveraging the optimal transport theory to mitigate the effects of high-variance particles. Additionally, exploiting the utilisation of irreversible Bayesian samplers to improve the mixing properties and convergence speed of Bayesian plastic tracker algorithms can be listed as another important research direction. 

Detecting where marine debris and floating plastics are located and tracking where they are heading are of crucial importance, however, there is one more important aspect about this: \textbf{the detection of the sources of the plastic.} Detecting and acting to clean plastic pollution from the oceans is an important action but requires repetitive efforts as soon as we do not stop the source of the pollution. Hence, the effort by Lavender \cite{lavender2022detection} via proposing a date set for plastic pollutants both on the land and coastal areas can be seen as one of the starting points of this effort since analyzing the source of the pollution requires discriminating the pollutants not only on land but also in small inland waters such as rivers. Similarly, the work by Sasaki et. al. \cite{Sasaki20226391} exploiting coastal debris detection via machine learning approaches and utilising Maxar WV-3 imagery accommodates the same importance in order to clearly discriminate the debris on the sea surface and on Land. 

As mentioned in the above sections, floating plastics' trajectory in the ocean has totally been affected by surface signatures and currents. Combining SAR and optical imagery information with the ground-breaking statistical samplers for tracking approaches will make analysing ocean currents, and wave structures possible, which can lead to developing back-tracking approaches to predict the sources of plastic pollution. Lastly, In order to mitigate the drawbacks of small-sized plastic bits and relatively low spatial resolution remote sensing data, group (or namely cloud) tracking-based approaches \cite{mihaylova2014overview,septier2009tracking} are of great importance.

\section{Conclusions}\label{sec:conc}
The ocean and waters of the planet Earth have been facing various threads during the last couple of decades. Whilst Global warming-related problems and some power-holders ignorance about it are still there, the Earth also dealing with pollution problems in the ocean, air, etc. Along with the technological developments, especially in the machine learning and artificial intelligence area in the last decade, computer vision approaches started to play some crucial roles to guide initiatives and decision-makers to act for the aforementioned problems. Marine debris and floating plastics are one of those problems, and remote sensing computational imaging efforts can help us to hear the call of the ocean!

This paper via presenting a review of the academic efforts in the literature aims to highlight developed computational remote sensing imaging approaches for the purpose of detecting marine debris and floating plastics. After carefully discussing each favourably valuable academic outcome, we also highlighted the literature statistics for marine debris monitoring purposes in the last decade. Finally, depending on the author's experience and research on marine-related remote sensing computational imaging approaches, various challenges, and potential future research directions have been listed (non-exhaustive).







\bibliographystyle{elsarticle-num} 
\bibliography{main}
\end{document}